\documentclass{pasj00}
\begin{document}
\SetRunningHead{J.\ Fukue and C.\ Akizuki}
{Radiative Transfer and Limb Darkening of Accretion Disks}
\Received{yyyy/mm/dd}
\Accepted{yyyy/mm/dd}

\title{Radiative Transfer and Limb Darkening of Accretion Disks}

\author{Jun \textsc{Fukue} and Chizuru \textsc{Akizuki}} 
\affil{Astronomical Institute, Osaka Kyoiku University, 
Asahigaoka, Kashiwara, Osaka 582-8582}
\email{fukue@cc.osaka-kyoiku.ac.jp}



\maketitle


\begin{abstract}
Transfer equation in a geometrically thin accretion disk 
is reexamined under the plane-parallel approximation
with finite optical depth.
Emergent intensity is analytically obtained
in the cases with or without internal heating.
For large or infinite optical depth,
the emergent intensity exhibits a usual limb-darkening effect,
where the intensity linearly changes
as a function of the direction cosine.
For small optical depth, on the other hand,
the angle-dependence of the emergent intensity 
drastically changes.
In the case without heating but with uniform incident radiation
at the disk equator,
the emergent intensity becomes isotropic for small optical depth.
In the case with uniform internal heating,
the limb brightening takes place for small optical depth.
We also emphasize and discuss the limb-darkening effect
in an accretion disk for several cases.
\end{abstract}

\section{Introduction}

Accretion disks are now widely believed
to be energy sources in various active phenomena in the universe:
in protoplanetary nebulae around young stellar objects,
in cataclysmic variables and supersoft X-ray sources,
in galactic X-ray binaries and microquasars, and
in active galaxies and quasars.
Accretion-disk models have been extensively studies
during these three decades (see Kato et al. 1998 for a review).

Radiative transfer in the accretion disk has been investigated
in relation to the structure of a static disk atmosphere
and the spectral energy distribution from the disk surface
(e.g., Meyer, Meyer-Hofmeister 1982; Cannizzo, Wheeler 1984;
K\v ri\v z and Hubeny 1986; Shaviv, Wehrse 1986; Adam et al. 1988;
Hubeny 1990; Ross et al. 1992; Artemova et al. 1996;
Hubeny, Hubeny 1997, 1998; Hubeny et al. 2000, 2001;
Davis et al. 2005; Hui et al. 2005).
In many cases the diffusion approximation or Eddington one
was employed;
it provides a satisfactory description at large optical depth,
although the emergent radiation field originates
at optical depth of the order of unity.
Furthermore, gray and non-gray models of accretion disks 
were constructed under numerical treatments
(K\v ri\v z and Hubeny 1986; Shaviv and Wehrse 1986;
Adam et al. 1988; Ross et al. 1992; Shimura and Takahara 1993;
Hubeny, Hubeny 1997, 1998; Hubeny et al. 2000, 2001;
Davis et al. 2005; Hui et al. 2005)
and under analytical ones
(Hubeny 1990; Artemova et al. 1996).

In these studies, however,
the vertical movement and the mass loss were not considered.
Hence, recently, in relation to the radiative disk wind,
radiative transfer in a moving disk atmosphere
was also investigated
(e.g., Fukue 2005a, b, 2006a, b).
In contrast to the static atmosphere,
in the moving atmosphere
the boundary condition at the surface of zero optical depth
should be modified (Fukue 2005a, b).
Moreover,
the usual Eddington approximation violates
in the highly relativistic flow (Fukue 2005b;
see also 
Turolla, Nobili 1988; Turolla et al. 1995; Dullemond 1999),
and the velocity-dependent variable Eddington factor
was proposed (Fukue 2006b).

In the usual studies of radiative transfer in the disk,
the emergent intensity has not been fully obtained,
since the attention were usually focused
on the disk internal structure
such as a temperature distribution.
In addition, the effect of limb darkening
has not been well examined
except for a few cases 
relating to cataclysmic variables 
(e.g., Diaz et al. 1996; Wade, Hubeny 1998;
see also Fukue 2000; Hui et al. 2005
for high energy cases).

In this paper,
we thus reexamine radiative transfer in the accretion disk
with {\it finite optical depth}
under the plane-parallel approximation,
and analytically obtain an emergent intensity
for the cases with or without internal heating.
Besides cataclysmic variables,
we also emphasize and discuss the limb-darkening effect
in an accretion disk for various cases.

In the next section
we describe the basic equations.
In section 3, we show analytical solutions.
In section 4,
we discuss several cases of accretion disk models,
and emphasize the importance of the limb-darkening effect.
The final section is devoted to concluding remarks.


\section{Basic Equations}

We here assume the followings:
(i) The disk is steady and axisymmetric.
(ii) It is also geometrically thin and plane parallel.
(iii) As a closure relation, we use the Eddington approximation.
(iv) The gray approximation, where the opacity does not depend on
frequency, is adopted.
(v) The viscous heating rate is concentrated at the equator
or uniform in the vertical direction.

The radiative transfer equations 
are given in several literatures
(Chandrasekhar 1960; Mihalas 1970; Rybicki, Lightman 1979;
Mihalas, Mihalas 1984; Shu 1991; Kato et al. 1998).
For the plane-parallel geometry in the vertical direction ($z$),
the frequency-integrated transfer equation,
the zeroth moment equation, and
the first moment equation become, respectively,
\begin{eqnarray}
   \cos \theta \frac{dI}{dz} & = & \rho \left[ \frac{j}{4\pi}
       -\left( \kappa_{\rm abs}+\kappa_{\rm sca} \right) I 
       + \kappa_{\rm sca} \frac{c}{4\pi}E \right],
\label{328_i.rad}
\\
   \frac{dF}{dz} & = & \rho \left( j - c\kappa_{\rm abs} E \right),
\label{328_f.rad}
\\
   \frac{dP}{dz} & = & - \frac{\rho (\kappa_{\rm abs}+\kappa_{\rm sca})}{c}F,
\label{328_p.rad}
\end{eqnarray}
where 
$\theta$ is the polar angle,
$I$ the frequency-integrated specific intensity,
$E$ the radiation energy density,
$F$ the vertical component of the radiative flux,
$P$ the $zz$-component of the radiation stress tensor, 
$\rho$ the gas density, and $c$ the speed of light.
The mass emissivity $j$ and opacity $\kappa_{\rm abs}$ and $\kappa_{\rm sca}$
are assumed to be independent of the frequency 
(gray approximation).

For matter, the vertical momentum balance
and energy equation are, respectively,
\begin{eqnarray}
   0 & = & -\frac{d\psi}{dz} - \frac{1}{\rho}\frac{dp}{dz}
                         + \frac{\kappa_{\rm abs}+\kappa_{\rm sca}}{c}F,
\label{328_hyd.rad}
\\
   0 & = & q^+_{\rm vis} - \rho \left( j - c\kappa_{\rm abs} E \right),
\label{328_energy.rad}
\end{eqnarray}
where $\psi$ is the gravitational potential,
$p$ the gas pressure, and
$q^+_{\rm vis}$ the viscous-heating rate.
In this paper, we do not solve the hydrostatic equilibrium (4).
Generally speaking, when the contribution of the radiative flux
is small, compared with the pressure gradient term,
the gas pressure dominates in the atmoshere, and
the density distribution will not be constant.
When the radiative flux is strong, on the other hand,
the radiation pressure dominates, and
the density may be approximately constant throughout
much of the disk.
Anyway, we suppose that the density distribution
would be ajusted so as to hold the hydrostatic equilibrium (4)
through the main part of the disk atmosphere,
under the radiative flux obtained later.

Using this energy equation (\ref{328_energy.rad})
and introducing the optical depth, defined by
\begin{equation}
   d\tau \equiv -\rho \left(\kappa_{\rm abs}+\kappa_{\rm sca}\right) dz,
\label{328_tau.rad}
\end{equation}
we rewrite the radiative transfer equations:
\begin{eqnarray}
   \mu \frac{dI}{d\tau} & = &  I - \frac{c}{4\pi}E
           -\frac{1}{4\pi} \frac{1}{\kappa_{\rm abs}+\kappa_{\rm sca}}
            \frac{q^+_{\rm vis}}{\rho},
\label{328_itau.rad}
\\
   \frac{dF}{d\tau} & = & -\frac{1}{\kappa_{\rm abs}+\kappa_{\rm sca}}
            \frac{q^+_{\rm vis}}{\rho},
\label{328_ftau.rad}
\\
   c\frac{dP}{d\tau} & = & F,
\label{328_ptau.rad}
\\
   cP &=& \frac{1}{3} cE,
\label{328_e.rad}
\end{eqnarray}
where $\mu \equiv \cos\theta$.
Final equation is the usual Eddington approximation.

As for the boundary condition at the disk surface of $\tau=0$,
we impose a usual condition:
\begin{equation}
   3cP_{\rm s} = cE_{\rm s} = 2F_{\rm s} ~~~~~{\rm at}~~\tau=0,
\end{equation} 
where the subscript s denotes the values at the disk surface.

For the internal heating,
we consider two extreme cases:
(i) No heating ($q^+_{\rm vis}=0$),
where the viscous heating is concentrated at the disk equator
and there is no heating source in the atmosphere.
(ii) Uniform heating in the sense that
$q^+_{\rm vis}/(\kappa^{\rm abs}+\kappa^{\rm sca})\rho = $constant.
The latter case means that the kinematic viscosity $\nu$
is constant in the vertical direction,
since $q^+_{\rm vis}/\rho = \nu (rd\Omega/dr)^2$,
as long as the opacities are constant.

Finally, the disk total optical depth becomes
\begin{equation}
   \tau_0 = - \int_H^0 \rho (\kappa_{\rm abs}+\kappa_{\rm sca}) dz,
\end{equation}
where $H$ is the disk half-thickness.

\section{Analytical Solutions}

Except for the emergent intensity $I$,
several analytical expressions for moments 
as well as temperature distributions
were obtained by several researchers
(e.g., Laor, Netzer 1989; Hubeny et al. 2005; Artemova et al. 1996).
For the completeness,
we recalculate them as well as the intensity $I$.

\subsection{No Heating Case}

We first consider the case without heating
in the disk atmosphere: $q^+_{\rm vis}=0$,
but with uniform incident intensity $I_0$ from the disk equator,
where the viscous heating is assumed to be concentrated.

In this case, the analytical solutions of moment equations
are easily given as
\begin{eqnarray}
   F &=& F_{\rm s} = \pi I_0,
\\
   3cP &=& cE = 3F_{\rm s} \left( \frac{2}{3} + \tau \right).
\end{eqnarray}
This is  a familar solution
under the Milne-Eddington approximation
for a plane-parallel geometry.
It should be noted that the vertical radiative flux $F$ is conserved,
and equals to $\pi I_0$ at the disk equator.

Since we obtain the radiation energy density $E$
in the explicit form,
we can now integrate the radiative transfer equation (\ref{328_itau.rad}).
After several partial integrations,
we obtain both an outward intensity $I(\tau, \mu)$ ($\mu>0$)
and an inward intensity $I(\tau, -\mu)$ as
\begin{eqnarray}
   I(\tau, \mu) &=& \frac{3F_{\rm s}}{4\pi}
         \left[
           \frac{2}{3} + \tau + \mu  
           -\left( \frac{2}{3} + \tau_0 + \mu
                    \right) e^{(\tau -\tau_0)/\mu}
            \right]
\nonumber \\
     && + I(\tau_0, \mu) e^{(\tau -\tau_0)/\mu},
\label{no_isol11.rad} \\
   I(\tau, -\mu) &=& \frac{3F_{\rm s}}{4\pi}
         \left[
           \frac{2}{3} + \tau - \mu  
           -\left( \frac{2}{3} - \mu
                    \right) e^{-\tau/\mu}
            \right],
\label{no_isol12.rad}
\end{eqnarray}
where $I(\tau_0, \mu)$ is the boundary value
at the midplane of the disk.

In the geometrically thin disk with finite optical depth $\tau_0$
and uniform incident intensity $I_0$ from the disk equator,
the boundary value $I(\tau_0, \mu)$ of the outward intensity $I$
consists of two parts:
\begin{equation}
   I(\tau_0, \mu) = I_0 + I(\tau_0, -\mu),
\end{equation}
where $I_0$ is the uniform incident intensity and
$I(\tau_0, -\mu)$ is the {\it inward} intensity from
the backside of the disk beyond the midplane.
Determining $I(\tau_0, -\mu)$ from equation (\ref{no_isol12.rad}),
we finally obtain the outward intensity as
\begin{eqnarray}
   I(\tau, \mu) &=& \frac{3F_{\rm s}}{4\pi}
         \left[
           \frac{2}{3} + \tau + \mu  -2\mu e^{(\tau -\tau_0)/\mu}
         \right.
\nonumber \\
   &&    \left.
           -\left( \frac{2}{3} - \mu \right) e^{(\tau -2\tau_0)/\mu}
            \right]
      + I_0 e^{(\tau -\tau_0)/\mu}
\nonumber \\
    &=& \frac{3F_{\rm s}}{4\pi}
         \left[
           \frac{2}{3} + \tau + \mu  
           +\left( \frac{4}{3} -2\mu \right) e^{(\tau -\tau_0)/\mu}
         \right.
\nonumber \\
   &&     \left.
           -\left( \frac{2}{3} - \mu \right) e^{(\tau -2\tau_0)/\mu}
            \right],
\label{no_isol13.rad}
\end{eqnarray}
where we have used $F_{\rm s}=\pi I_0$.

For sufficiently large optical depth $\tau_0$,
this equation (\ref{no_isol13.rad}) reduces to
the usual Milne-Eddington solution:
\begin{equation}
   I = \frac{3F_{\rm s}}{4\pi} \left( \frac{2}{3} + \tau + \mu \right).
\end{equation}

Finally, the emergent intensity $I(0, \mu)$ emitted from the disk surface
for the finite optical depth becomes
\begin{eqnarray}
   I(0, \mu) &=& \frac{3F_{\rm s}}{4\pi}
         \left[ \frac{2}{3} + \mu 
           +\left( \frac{4}{3} -2\mu \right) e^{-\tau_0/\mu}
         \right.
\nonumber \\
    &&     \left.
            -\left( \frac{2}{3} -\mu \right) e^{-2\tau_0/\mu}
            \right].
\label{no_isol0.rad}
\end{eqnarray}

\begin{figure}
  \begin{center}
  \FigureFile(80mm,80mm){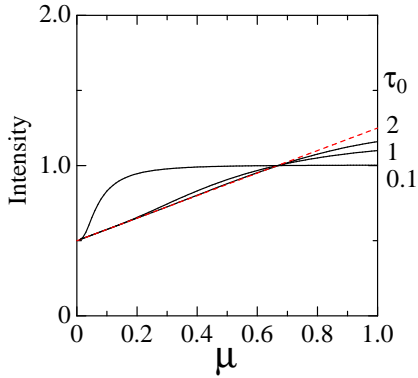}
  \end{center}
\caption{
Normalized emergent intensity
as a function of $\mu$
for the case without heating.
The numbers attached on each curve
are values of $\tau_0$ at the disk midplane.
The dashed straight line is for the usual
plane-parallel case with infinite optical depth.
}
\end{figure}

In figure 1,
the emergent intensity $I(0, \mu)$
normalized by the isotropic value $\bar{I}$ ($=F_{\rm s}/\pi$)
is shown for several values of $\tau_0$
as a function of $\mu$.

As is easily seen in figure 1,
for large optical depth ($\tau_0 >10$)
the angle-dependence of the emergent intensity
is very close to the case for a usual plane-parallel case
with infinite optical depth.
Therefore, the usual limb-darkening effect is seen.
Namely, in the case of a semi-infinite disk with large optical depth,
the energy density increases linearly with the optical depth
in the atmosphere, and the temperature increases accordingly.
As a result, an observer at a pole-on position of $\mu=1$
will see deeper in the disk, where the temperature
(and therefore, the source function) is larger than
that observed by an observer at an edge-on position of $\mu=0$.
Thus, the observed intensity will be higher at $\mu=1$.
This is just a usual limb-darkening.

For small optical depth, however,
the angle-dependence is drastically changed.
When the optical depth is a few,
the vertical intensity ($\mu \sim 1$) decreases
due to the finiteness of the optical depth.
That is, we cannot see the `deeper' position in the atmosphere,
compared with the case of a semi-infinite disk.
Furthermore, when the optical depth is less than unity,
the intensity in the direction of small $\mu$ increases,
and the emergent intensity becomes isotropic
with a uniform value $I_0$ at the disk equator;
the limb-darkening effect disappears.
Indeed, the limiting case of $\tau_0 \sim 0$,
$I(0, \mu) \sim F_{\rm s}/\pi$.
That is, in this case of very small optical depth,
the source function is dominated
by the isotropic source at the midplane.

\subsection{Uniform Heating Case}

Now, we consider the case with uniform heating:
$q^+_{\rm vis}/(\kappa_{\rm abs}+\kappa_{\rm sca})\rho = $constant.

Integrating the equation (\ref{328_ftau.rad})
 under the following boundary conditions:
\begin{eqnarray}
   F = 0         & \quad {\rm at} \quad & \tau = \tau_0, \nonumber \\
   F = F_{\rm s} & \quad {\rm at} \quad & \tau = 0,
\end{eqnarray}
we obtain
\begin{equation}
   F = F_{\rm s} \left( 1 - \frac{\tau}{\tau_0} \right).
\label{328_fsol.rad}
\end{equation}
The radiative flux $F$ linearly increases from 0 
to the surface value $F_{\rm s}$.

Substituting equation (\ref{328_fsol.rad}) into equation (\ref{328_ptau.rad}),
and integrating the resultant equation under the boundary condition (11),
we obtain
\begin{equation}
   3cP = cE = 3F_{\rm s} \left( \frac{2}{3} + \tau
                         -\frac{\tau^2}{2\tau_0} \right).
\label{328_psol.rad}
\end{equation}
This expression (\ref{328_psol.rad}) for finite optical depth
is seen in, e.g., Laor and Netzer (1989).
A similar but more general expression was obtained
by Hubeny (1990).
In any case,
this expression reduces to the Milne-Eddington solution
for sufficiently large optical depth.
In the case of finite optical depth,
the radiation energy density and pressure decrease from the midplane
to the surface in the quadratic form.
It should be noted that
at the midplane of the disk of $\tau=\tau_0$,
\begin{equation}
   3cP = cE = 3F_{\rm s} \left( \frac{2}{3} + \frac{\tau_0}{2} \right).
\label{328_psol2.rad}
\end{equation}
As already mentioned by Hubeny (1990),
the energy density at the disk midplane is
the half of the corresponding stellar atmospheric one.
This is explained by the fact that
the radiation from the disk midplane may escape equally
to both sides of the disk.

Since we obtain the radiation energy density $E$
in the explicit form (\ref{328_psol.rad}),
we can now integrate the radiative transfer equation (\ref{328_itau.rad}).
After several partial integrations,
we obtain both an outward intensity $I(\tau, \mu)$ ($\mu>0$)
and an inward intensity $I(\tau, -\mu)$ as
\begin{eqnarray}
   I(\tau, \mu) &=& \frac{3F_{\rm s}}{4\pi}
         \left[
           \frac{2}{3} + \tau + \mu  
           +\frac{1}{\tau_0} \left( \frac{1}{3} -\frac{\tau^2}{2}
                            -\mu \tau -\mu^2 \right)
         \right.
\nonumber \\
    &&   \left.
           -\left( \frac{2}{3} + \frac{\tau_0}{2} + \frac{1}{3\tau_0}
                   -\frac{\mu^2}{\tau_0} \right) e^{(\tau -\tau_0)/\mu}
            \right]
\nonumber \\
    &&      + I(\tau_0, \mu) e^{(\tau -\tau_0)/\mu},
\label{heat_isol11.rad} \\
   I(\tau, -\mu) &=& \frac{3F_{\rm s}}{4\pi}
         \left[
           \frac{2}{3} + \tau - \mu  
           +\frac{1}{\tau_0} \left( \frac{1}{3} -\frac{\tau^2}{2}
                             +\mu \tau -\mu^2 \right)
         \right.
\nonumber \\
       &&  \left.
           -\left( \frac{2}{3} - \mu + \frac{1}{3\tau_0}
                   -\frac{\mu^2}{\tau_0} \right) e^{-\tau/\mu}
            \right],
\label{heat_isol12.rad}
\end{eqnarray}
where $I(\tau_0, \mu)$ is the boundary value
at the midplane of the disk.

In the case with uniform heating and without the incident intensity,
the boundary value $I(\tau_0, \mu)$ of the outward intensity $I$ is
\begin{equation}
   I(\tau_0, \mu) =  I(\tau_0, -\mu),
\end{equation}
and we finally obtain the outward intensity as
\begin{eqnarray}
   I(\tau, \mu) &=& \frac{3F_{\rm s}}{4\pi}
         \left[
           \frac{2}{3} + \tau + \mu  
           +\frac{1}{\tau_0} \left( \frac{1}{3} -\frac{\tau^2}{2}
                             -\mu \tau -\mu^2 \right)
         \right.
\nonumber \\
   &&    \left.
           -\left( \frac{2}{3} - \mu +\frac{1}{3\tau_0}
                   -\frac{\mu^2}{\tau_0} \right) e^{(\tau -2\tau_0)/\mu}.
            \right]
\label{heat_isol.rad}
\end{eqnarray}

For sufficiently large optical depth $\tau_0$,
this equation (\ref{heat_isol.rad}) also reduces to
the usual Milne-Eddington solution (19).

Finally, the emergent intensity $I(0, \mu)$ emitted from the disk surface
for the finite optical depth becomes
\begin{eqnarray}
   I(0, \mu) &=& \frac{3F_{\rm s}}{4\pi}
         \left[ \frac{2}{3} + \mu
           +\frac{1}{\tau_0} \left( \frac{1}{3} -\mu^2 \right)
         \right.
\nonumber \\
     & & \left.
           -\left( \frac{2}{3} -\mu + \frac{1}{3\tau_0}
                   -\frac{\mu^2}{\tau_0} \right) e^{-2\tau_0/\mu}
            \right].
\label{328_isol0.rad}
\end{eqnarray}

\begin{figure}
  \begin{center}
  \FigureFile(80mm,80mm){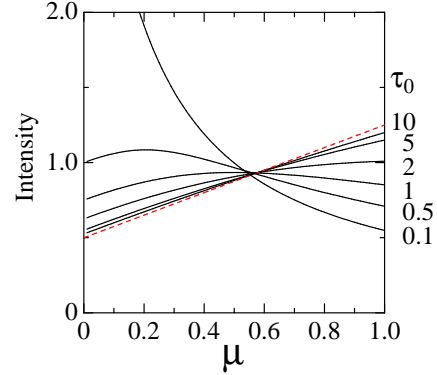}
  \end{center}
\caption{
Normalized emergent intensity
as a function of $\mu$
for the case with uniform heating.
The numbers attached on each curve
are values of $\tau_0$ at the disk midplane.
The dashed straight line is for the usual
plane-parallel case with infinite optical depth.
}
\end{figure}

In figure 2,
the emergent intensity $I(0, \mu)$
normalized by the isotropic value $\bar{I}$ ($=F_{\rm s}/\pi$)
is shown for several values of $\tau_0$
as a function of $\mu$.

As is easily seen in figure 2,
for large optical depth ($\tau_0 >10$)
the angle-dependence of the emergent intensity
is very close to the case with a usual plane-parallel case
with infinite optical depth.
Therefore, the usual limb-darkening effect is seen,
as already stated at the end of section 3.1.

For small optical depth, however,
the angle-dependence is drastically changed
similar to the case without heating.
In the vertical direction of $\mu \sim 1$,
the emergent intensity decreases as the optical depth decreases.
This is due to the finiteness of the optical depth.
That is, we cannot see the `deeper' position in the atmosphere,
compared with the case of a semi-infinite disk.
In the inclined direction of small $\mu$, on the other hand,
the emergent intensity becomes larger than that
in the case of the infinite optical depth.
Moreover, when the optical depth is less than unity,
the emergent intensity for small $\mu$ is greater than unity:
the {\it limb brightening} takes place.
Indeed, in the limiting case of $\tau_0 \sim 0$,
$I(0, \mu) \sim (F_{\rm s}/\pi)/(2\mu)$.
This is because that the path length is longer
for such a case of small $\mu$.
That is, in this case for low optical depth,
the source function is very uniform.
This, coupled with the absence of an isotropic source at the midplane,
is why the geometric effect (longer path length)
is dominant and one finds limb `brightening'.

\subsection{Validity of the Eddington Approximation}

In this subsection, we briefly discuss the validity
of the closure relation in the present treatment.
In this paper, we have adopted the usual Eddington approximation,
where the ratio of the radiation pressure to the energy density
is fixed as 1/3, to close the moment equations.
As is well known, this approximation is correct
in the limit of an isotropic radiation field.
Hence, in the problem of limb-darkening,
where the deviation from the isotropy is essential,
this approximation is only approximately correct,
although it is used in the usual Milne-Eddington approximation.

For example, let us suppose the case of a semi-infinite disk
with an infinite optical depth $\tau_0$,
In this case, we easily calculate
the energy density as well as the radiation pressure
 from the derived intensity (19) of the Milne-Eddington solution.
At the deeper position in the atmosphere,
where the integration is done in all directions,
the re-calculated variables satisfy the condition of
$P/E=1/3$.
However, at the surface of the disk, 
where the integration should be done in a semi-sphere,
this is not true.
At the surface of $\tau=0$, we integrate the emergent intensity
to yield:
\begin{eqnarray}
   cE &=& \frac{3F_{\rm s}}{2} \int_0^1
          \left( \frac{2}{3}+\mu \right) d\mu
       = 3F_{\rm s} \frac{7}{12},
\\
   cP &=& \frac{3F_{\rm s}}{2} \int_0^1
          \left( \frac{2}{3}+\mu \right)\mu^2 d\mu
       = \frac{1}{3} 3F_{\rm s} \frac{17}{24},
\end{eqnarray}
or $P/E=(1/3)(17/14)$.
Hence, in the case under the usual limb-darkening effect,
the ratio of the radiation pressure to the energy density
is slightly larger than 1/3 at the surface.
This is just the peaking effect
originated from the anisotropic radiation field
near to the disk surface.

On the other hand, in the limb-brightening case
of small optical depth with uniform heating,
the situation is reversed.
In this case,
the ratio of the radiation pressure to the energy density
becomes smaller than 1/3,
and the Eddington approximation holds approximately.
This is also originated from the anisotropy,
and may be called an {\it anti-peaking} effect.
As is seen in figure 2,
the limb-brightening becomes stronger and stronger
for small optical depth.
Hence, for such a case of very small optical depth,
the Eddington approximation would not be good,
although the qualitative properties would not be changed.

In order to obtain the intensity distribution more precisely,
we, for example, introduce a variable Eddington factor,
that is beyond the scope of the present paper.

\section{Discussion}

As was derived in the previous section,
the emergent intensity $I$ of the accretion disk
depends on the disk total optical depth $\tau_0$
as well as the direction cosine $\mu$.
The limb-darkening effect is considerably modified
for small $\tau_0$,
compared with the usual case for infinite optical depth.
Even for the case with sufficiently large optical depth,
limb darkening in the luminous accretion disk
must be important, and should be examined more carefully.

In this section,
we discuss several cases in turn,
and call the attention to the importance
of the limb-darkening effect.

\subsection{Standard Disk}

The optical depth at the midplane
in the inner region of a geometrically thin
standard disk (Shakura, Sunyaev 1973;
see also Kato et al. 1998)
is expressed as
\begin{equation}
    \tau_0 = \frac{1}{2}\kappa \Sigma
           = 20 \alpha^{-1}{\dot{m}}^{-1} {\hat{r}}^{3/2}
                \left( 1- \sqrt{ \frac{3}{\hat{r}} } \right)^{-1},
\end{equation}
where $\kappa$ is the electron scattering optical depth,
$\alpha$ the viscous parameter,
$\dot{m}$ the mass accretion rate normalized
by the critical rate $\dot{M}$ ($=L_{\rm E}/c^2$),
$L_{\rm E}$ being the Eddington luminosity of the central object,
$\hat{r}$ the radius normalized
by the Schwarzschild radius $r_{\rm g}$ ($=2GM/c^2$).

This optical depth becomes small
for large $\dot{m}$ and/or small $\hat{r}$.
For a slightly large accretion rate,
inside some critical radius
\begin{equation}
     r_{\rm cr} = 2\dot{m},
\end{equation}
the disk shifts to a supercritical regime,
whereas the disk is a standard regime outside $r_{\rm cr}$ (Fukue 2004).
At this critical radius,
the optical depth becomes
\begin{equation}
    \tau_{\rm cr} \sim 57 \alpha^{-1} \dot{m}^{1/2}.
\end{equation}
Hence, in a usual situation
the optical depth of the inner region of the standard disk
is greater than  several tens.

In such a situation, however,
due to a usual limb-darkening effect
for semi-infinite medium,
the emergent radiation toward the pole-on direction
is enchanced by 20 percent,
while
the emergent radiation seen from the edge-on direction
diminishes by 50 percent.
Thus,
in calculating the flux and spectrum of the standard disk,
we carefully consider the limb-darkening effect.

In the region inside the inner edge at $3r_{\rm g}$,
the disk gas freely falls toward the central black hole,
and the surface density (i.e., the disk optical depth)
quickly drops.
Hence, the emergent spectrum from the innermost region
inside the inner edge
would be greatly modified from the optically thick case
and the optically thin one.

\subsection{Supercritical Disk}

In the supercritical accretion disk,
where the mass accretion rate exceeds the critical rate,
the expression for the disk optical depth is changed.
For example, in the self-similar model without mass loss
(Fukue 2000),
the optical depth at the midplane of the disk is
\begin{equation}
   \tau_0 = \frac{\kappa}{4\pi} \frac{\dot{M}}{c_1 \alpha}
            \frac{1}{\sqrt{GMr}}
          = \frac{\dot{m}}{\sqrt{2} c_1 \alpha} \frac{1}{\hat{r}},
\end{equation}
where $c_1$ is a coefficient of the order of unity.
At the critical radius, the disk optical depth is
\begin{equation}
   \tau_{\rm cr} \sim \frac{1}{2c_1 \alpha} \dot{m}^{1/2}.
\end{equation}

In the critical accretion disk (Fukue 2004),
where the mass accretion rate exceeds the critical one,
but the excess mass is expelled by the wind mass loss,
the optical depth at the midplane of the disk is 
\begin{equation}
    \tau_0 = \frac{16\sqrt{6}}{\alpha} {\hat{r}}^{1/2}
           = 39.2 \alpha^{-1} \hat{r}^{1/2}.
\end{equation}
At the critical radius, the disk optical depth is
\begin{equation}
   \tau_{\rm cr} \sim 55 \alpha^{-1} \dot{m}^{1/2}.
\end{equation}

Hence, in a usual situation
the optical depth of the supercritical/critical disk
is also greater than  several tens.

In such a situation, however,
the usual limb-darkening effect
for semi-infinite medium is also important (see Fukue 2000).
In addition to the limb-darkening effect,
the geometrical effects,
such as a {\it projection effect} and a {\it self-occultation},
should be considered
in calculating the flux and spectrum of the supercritical disk
(e.g., Fukue 2000; Watarai et al. 2005; Kawata et al. 2006).

\subsection{Disk Corona and ADAF}

If a luminous disk is sandwiched by a disk corona,
the situation is similar to the case
without heating, but with the incident intensity from the midplane.
Hence,
when the optical depth of the corona is sufficiently smaller than unity,
the emergent intensity is just that of the disk,
except for a very small direction cosine.
However,
when the optical depth of the corona is a few,
limb darkening takes place,
and the emergent intensity toward the edge-on direction
remarkably reduces.

On the other hand,
if the accretion rate is quite small, and
the inner region of the disk becomes
an optically-thin advection dominated state (ADAF),
the situation is similar to the case with internal heating,
although, rigorously speaking,
the plane-parallel approximation may be invalid.
When the optical depth of an ADAF region is sufficiently smaller than unity,
the limb brightening would take place.

In both cases with hot gas, however,
the effect of the Compton scattering may be important.
Hence, the angle dependence of the intensity would be
modified by the angle dependence of the Compton scattering,
and the transfer problem should be treated more carefully.

In addition, in the latter case of ADAF,
the accretion flow is supposed to be conical or spherical.
Hence, when the opening angle is small,
the limb brightening would qualitatively take place.
When the opening angle is large, on the other hand,
the extension of the emitting region is large,
and the angle dependence of the emergent intensity
would become much more complicated
than the present simple case.

\subsection{Relativistic Disk}

In the case of the relativistic standard disk
(e.g., Novikov, Thorne 1973; Page, Thorne 1974),
the situation is similar to the non-relativistic case.
However,
the direction cosine of the local emergent intensity
in the disk to the observer is changed by two additional reasons:
(i) the light trajectory is bent by the space-time curvature, and
(ii) the emission from the gas rotating around a black hole
suffers from the special relativistic aberration.
Limb-darkening effect in the comoving frame
on the spectral energy distribution (SED) from
the relativistic accretion disk was considered
by, e.g., Fu and Taam (1990) and Gierli\'nski et al. (2001).

Recently, due to submilliarcsecond astrometry,
imaging of a black-hole {\it silhouette}
is expected to become possible in the near future
at infrared and submillimetre wavelengths.
The ``photographs'' of relativistic accretion disks
around a black hole have been obtained 
by many researchers
(Luminet 1979; Fukue, Yokoyama 1988;
Karas et al. 1992; Jaroszy\'nski et al. 1992;
Fanton et al. 1997; Fukue 2003; Takahashi 2004, 2005).
In these studies, however,
the usual limb darkening was not considered.

\begin{figure}
  \begin{center}
  \FigureFile(80mm,80mm){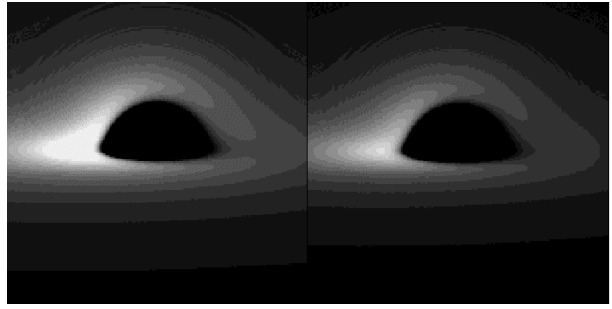}
  \end{center}
\caption{
Edge-on view of a dressed black hole
without (left panel) and with (right panel) limb darkening.
The inclination angle is 80$^\circ$.
}
\end{figure}

\begin{figure}
  \begin{center}
  \FigureFile(80mm,80mm){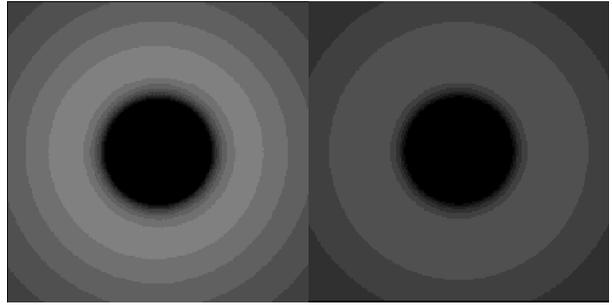}
  \end{center}
\caption{
Pole-on view of a dressed black hole
without (left panel) and with (right panel) limb darkening.
The inclination angle is 0$^\circ$.
}
\end{figure}

Examples of silhouettes of a dressed black hole
are shown in figures 3 and 4.
In figure 3 an edge-on view with inclination angle of 80$^\circ$ is shown,
while a pole-on view is expressed in figure 4.
In both figures,
the left panels are for the case without limb darkening,
whereas the right panels are for the case with limb darkening.

In the edge-on view (figure 3),
the image of a limb-darkening disk darkens as expected.
This is due mainly to the usual limb-darkening effect
for small direction cosine.
Surprisingly, on the other hand,
in the pole-on view (figure 4),
the image of a limb-darkeing disk also darkens!
This is because that
the local direction cosine becomes small
due to the light aberration associated with the disk rotation.

Thus, 
in calculating spectra and observed fluxed of relativistic disks
as well as in taking black hole silhouettes,
we should carefully consider limb darkening.
For example, 
besides Fu and Taam (1990) and Gierli\'nski et al. (2001),
Hui et al. (2005) discussed limb darkening in some details
in their non-LTE calculation of accretion disk spectra
around intermediate-mass black holes.


\section{Concluding Remarks}

In this paper
we analytically solve the radiative transfer problem
of a geometrically thin accretion disk
with finite optical depth, and
obtain analytical expressions
for emergent intensity from the surface of the disk.
For small optical depth, 
the angle-dependence of the emergent intensity 
drastically changes from the case with large optical depth.
In the case without heating but with uniform incident radiation
at the disk equator,
the emergent intensity becomes isotropic for small optical depth.
In the case with uniform internal heating,
the limb brightening takes place for small optical depth.
We also emphasize
the importance of limb darkening 
in the accretion disk study, and discuss
several cases, including relativistic silhouettes.

In the calculation of spectra from, e.g., cataclysmic variables,
the effect of limb darkening was considered
(e.g., Diaz et al. 1996; Wade, Hubeny 1998).
In the cases of high energy and relativistic regimes,
we should also consider limb darkening
(cf. Hui et al. 2005).

In addition,
if there exist intense radiation sources,
such as neutron stars or radiating jets,
irradiation takes place and
the outer boundary condition changes (e.g., Hubeny 1990).
In such cases under strong incident radiation,
limb brightening may occur
(cf. Stibbs 1971),
even for large optical depth.

Finally, if there exists mass loss from the disk surface,
the radiative transfer problem becomes much more complicated.
In order to examine spectra, fluxes, eclipsing light curves,
we must calculate the emergent intensity
analytically or numerically.

\vspace*{1pc}

This work has been supported in part
by a Grant-in-Aid for Scientific Research (18540240 J.F.) 
of the Ministry of Education, Culture, Sports, Science and Technology.


\end{document}